\documentclass[twocolumn,english]{revtex4-1}
\usepackage[T1]{fontenc}
\usepackage[latin9]{inputenc}
\usepackage{amsmath}
\usepackage{amssymb}
\usepackage{graphicx}


\usepackage{babel}
\begin{document}
\title{Coexistence of two kinds of superfluidity in Bose-Hubbard model with
density-induced tunneling at finite temperatures}
\author{A. Krzywicka, T. P. Polak}
\affiliation{Faculty of Physics, Adam Mickiewicz University, Pozna\'{n}, Poland}
\begin{abstract}
With use of the U(1) quantum rotor method in the path integral effective
action formulation, we have confirmed the mathematical similarity
of the phase Hamiltonian and of the extended Bose-Hubbard model with
density-induced tunneling (DIT). Moreover, we have shown that the
latter model can be mapped to a pseudospin Hamiltonian that exhibits
two coexisting (single-particle and pair) superfluid phases. Phase
separation of the two has also been confirmed, determining that there
exists a range of coefficients in which only pair condensation, and
not single-particle superfluidity, is present. The DIT part supports
the coherence in the system at high densities and low temperatures,
but also has dissipative effects independent of the system's thermal
properties.
\end{abstract}
\maketitle

\section{Introduction}

Optical lattices provide an excellent framework for studying many-body
Hamiltonians, which are difficult to replicate in solids due to their
complexity and lack of control over various parameters. The Hubbard
model, which contains a quantum phase transition between two ground
states: the superfluid state {[}\citealp{ketterle}{]} and the Mott
insulator state, is a staple of the study of strongly correlated systems
in low temperatures, and its various iterations have lately been under
particular scrutiny in relation to optical lattices (as well as Josephson
junction arrays in the bosonic case; see {[}\citealp{bruder}{]}).
{[}\citealp{duchon,greiner,jaksch}{]} Our interest lies in bosonic
systems, described by the Bose-Hubbard (BH) model and its extensions.
The BH model can be obtained as an approximation of a general second
quantization many-body Hamiltonian, describing a gas of interacting
bosons in an external potential, by cutting off all but the two most
important terms, ie. the ones that contribute most to the total energy:
the on-site two-particle interaction, $U$, and $t$, the single-particle
tunneling between two nearest-neighboring sites. {[}\citealp{bloch2008}{]}
Extended BH models are obtained by adding one or more of the cut interactions
to the pure BH model. Of course, this greatly increases the complexity
of the model, making exact analysis difficult -- and thus a comparatively
lacking section of condensed matter physics. Of all these interactions,
density-induced tunneling, known also as bond-charge interaction or
correlated hopping, contributes most to a system's energy {[}\citealp{luhmann,hirsch89,jurgensen2}{]},
and has successfully been experimentally observed on optical lattices
{[}\citealp{jurgensenexp,exp2}{]}, making it the most interesting
extension to work with.

In this work, we carry out a path integral analysis {[}\citealp{pathint}{]}
of the density-induced tunneling BH model, utilizing the U(1) quantum
rotor method {[}\citealp{refqr,kopec3}{]}, which replaces bosonic
field operators with interacting U(1) phase fields, leading to an
effective action formulation of the system's partition function. The
methods used allow us to make an explicit analytical connection between
this model and an extended Quantum Phase Model (QPM) Hamiltonian,
which describes pair tunnelling in bosonic many-body systems {[}\citealp{pairmf,pairkopec}{]},
thus showing that this behavior is anticipated by the bosonic density-induced
tunneling model, provided that many-body correlations are not excluded
from its analysis.

This model contains three-body correlations, as seen in the density-induced
tunneling term Eq. (\ref{eq:density}), which is a product of three
bosonic field operators. Previous considerations of this model {[}\citealp{density8,density11,density12,luhmann,jurgensen2}{]}
made use of mean field approximations, which do not account for such
correlations, as mean fields serve as replacements for any multi-linear
interactions. Thus, while the influence of density-induced tunneling
on the BH phase diagram, which describe the transitions between the
two ground states of the BH model -- Mott insulator and superfluid
-- which mean fields do retain, is well documented {[}\citealp{lewenstein}{]},
the presence of bosonic pairing has thus far remained analytically
unconfirmed at finite temperatures and beyond mean field level.

Furthermore, we map the newly-acquired phase Hamiltonian onto an $S=1$
pseudospin model {[}\citealp{refkopec,simanek}{]} and apply a mean
field approximation (which at this point does not erase the correlations
we wanted to preserve; the information has been absorbed into the
coefficients and the properties of the phase Hamiltonian), allowing
us to obtain phase diagrams via self-consistent critical line equations.
These temperature-dependent diagrams show the critical lines between
the normal phase and two others: the known single-particle superfluid
phase and the previously unconfirmed for density-induced tunneling
BH pair condensation at finite temperatures.

Here follows an outline of the contents of this publication. In Sec.
II, the model Hamiltonian is defined. In Sec. III, we introduce the
quantum rotor representation and derive an effective action for the
model. This effective action corresponds to the phase Hamiltonian.
Next, we map the obtained phase Hamiltonian onto $S=1$ pseudospin,
and calculate the critical line equations needed to analyze the thermodynamics
of the system. Exemplary diagrams of order parameters and specific
heat are shown and commented on in Sec. IV, followed by a summary
in Sec. V.

\section{Model Hamiltonian}

The Hamiltonian for this model consists of two parts:
\begin{align}
\mathcal{H} & =\mathcal{H}_{BH}+\mathcal{H}_{DIT},
\end{align}
where
\begin{align}
\mathcal{H}_{BH}= & \frac{U}{2}\sum_{i}n_{i}\left(n_{i}-1\right)-\sum_{\left\langle i,j\right\rangle }t_{ij}a_{i}^{\dagger}a_{j}-\mu\sum_{i}n_{i}
\end{align}
is the pure Bose-Hubbard model Hamiltonian, with $a_{i}^{\dagger}$,
$a_{i}$ being the bosonic creation and annihilation operators respectively,
obeying the canonical commutation relation $\left[a_{i},a_{j}^{\dagger}\right]=\delta_{ij}$,
and $n_{i}=a_{i}^{\dagger}a_{i}$ being the boson number operator
on site $i$. Further, $U>0$ is the on-site repulsion, $\mu$is the
chemical potential, $\left\langle i,j\right\rangle $ identifies a
summation over nearest neighbor sites, and $t_{ij}$ is the hopping
integral, the dispersion of which on a bipartite lattice in $d$ dimensions
is
\begin{equation}
t\left(\boldsymbol{k}\right)=2t\sum_{l=1}^{d}\cos k_{l}.
\end{equation}
This work focuses on the properties of the BH model with density-induced
tunneling on a simple cubic lattice. We also assume that hopping is
isotropic, $t_{ij}=t$. The density-induced tunneling (DIT) term is
\begin{equation}
\mathcal{H}_{DIT}=-T\sum_{\left\langle i,j\right\rangle }\left[a_{i}^{\dagger}\left(n_{i}+n_{j}\right)a_{j}+a_{j}^{\dagger}\left(n_{i}+n_{j}\right)a_{i}\right],\label{eq:density}
\end{equation}
with density-induced tunneling amplitude $T$. The full Hamiltonian
can be rewritten in a pure BH-like form:
\begin{equation}
\mathcal{H}=\frac{U}{2}\sum_{i}n_{i}^{2}-J\sum_{\left\langle i,j\right\rangle }a_{i}^{\dagger}a_{j}-\sum_{\left\langle i,j\right\rangle }\tilde{\mu}_{ij}n_{i},\label{eq:hamiltonian}
\end{equation}
with the coefficients
\begin{align}
J & =t-2T,\\
\tilde{\mu}_{ij} & =\bar{\mu}+4Ta_{i}^{\dagger}a_{j},\label{eq:chempot}\\
\bar{\mu} & =\frac{U}{2}+\mu-2T.
\end{align}
It is worth mentioning that the shifted chemical potential $\tilde{\mu}_{ij}$
is now an operator, due to the presence of the density-induced tunneling
amplitude $T$.

\section{Method}

\subsection{Quantum rotor approximation}

Using the quantum rotor method {[}\citealp{refqr}{]}, we will rewrite
the model as phase-only, and then carry out transformation to a pseudospin
model, much as in {[}\citealp{refkopec}{]}.

\subsubsection{Hubbard-Stratonovich and gauge transformations}

The path integral formulation of the partition function is
\begin{equation}
\mathcal{Z}=\int\left[\mathcal{D}\bar{a}\mathcal{D}a\right]e^{-\mathcal{S}\left[\bar{a},a\right]},
\end{equation}
where $\mathcal{S}$ is the effective action,
\begin{equation}
\mathcal{S}=\sum_{i}\int_{0}^{\beta}d\tau\,\bar{a}_{i}\left(\tau\right)\frac{\partial}{\partial\tau}a_{i}\left(\tau\right)+\int_{0}^{\beta}d\tau\,\mathcal{H}\left(\tau\right).\label{eq:action0}
\end{equation}
The bosonic field operators $a_{i}^{\dagger}$, $a_{i}$ are now represented
by complex fields $a_{i}\left(\tau\right)$ and $\mathcal{H}\left(\tau\right)=\mathcal{H}\left[\bar{a}\left(\tau\right),a\left(\tau\right)\right]$
is our Hamiltonian (Eq. (\ref{eq:hamiltonian})). Our first step is
decoupling the bilinear term in $\mathcal{H}$ by a Hubbard-Stratonovich
transformation, introducing the auxiliary fields $V_{i}\left(\tau\right)$:
\begin{equation}
e^{-\frac{U}{2}\sum_{i}\int d\tau\,n_{i}^{2}\left(\tau\right)}=\int\frac{dV}{2\pi}\,e^{-\sum_{i}\int d\tau\left[\frac{V_{i}^{2}\left(\tau\right)}{2U}-iV_{i}\left(\tau\right)n_{i}\left(\tau\right)\right]},
\end{equation}
which allows us to split the effective action (Eq. (\ref{eq:action0}))
into two terms, one of which is independent of the fields $a_{i}\left(\tau\right)$:
\begin{align}
\mathcal{Z} & =\int\left[\mathcal{D}\bar{a}\mathcal{D}a\right]e^{-\mathcal{S}_{1}\left[\bar{a},a\right]}\int\frac{dV}{2\pi}e^{-\mathcal{S}_{2}\left[n,V\right]},\label{eq:partf}\\
\mathcal{S}_{1} & =\int_{0}^{\beta}d\tau\left[\sum_{i}\bar{a}_{i}\left(\tau\right)\frac{\partial}{\partial\tau}a_{i}\left(\tau\right)-J\sum_{\left\langle i,j\right\rangle }\bar{a}_{i}\left(\tau\right)a_{j}\left(\tau\right)\right],\label{eq:s1}\\
\mathcal{S}_{2} & =\sum_{i}\int_{0}^{\beta}d\tau\left[\frac{1}{2U}V_{i}^{2}\left(\tau\right)-\left(iV_{i}\left(\tau\right)+\bar{\mu}\right)n_{i}\left(\tau\right)\right].
\end{align}
Next, we shift the electrochemical potential $V_{i}\left(\tau\right)=V_{i}^{T}\left(\tau\right)-\frac{\bar{\mu}}{i}$,
getting
\begin{align}
\mathcal{S}_{2}= & \sum_{i}\int_{0}^{\beta}d\tau\,\Biggl[\frac{1}{2U}\left(V_{i}^{T}\left(\tau\right)\right)^{2}-\frac{1}{2U}\bar{\mu}^{2}+\nonumber \\
 & -\frac{V_{i}^{T}\left(\tau\right)\bar{\mu}}{iU}-iV_{i}^{T}\left(\tau\right)n_{i}\left(\tau\right)\Biggr].
\end{align}
 $V_{i}^{T}$ is further split into static and periodic parts, 
\begin{equation}
V_{i}^{T}\left(\tau\right)=V_{i}^{S}\left(\tau\right)+V_{i}^{P}\left(\tau\right),
\end{equation}
which are defined as follows:
\begin{align}
V_{i}^{S}\left(\tau\right) & =\frac{1}{\beta}V_{i}^{T}\left(\omega_{m=0}\right),\\
V_{i}^{P}\left(\tau\right) & =\frac{1}{\beta}\sum_{m=1}^{+\infty}\left(V_{i}^{T}\left(\omega_{m}\right)e^{i\omega_{m}\tau}+c.c.\right),\label{eq:vperiodic}
\end{align}
where $\omega_{m}=2\pi m/\beta$ for integer values of $m$ are the
bosonic Matsubara frequencies. We then bind the periodic part of the
field $V_{i}^{P}\left(\tau\right)$ from Eq. (\ref{eq:vperiodic})
to a $U(1)$ phase field $\phi\left(\tau\right)$ via Josephson coupling:
\begin{equation}
V_{i}^{P}\left(\tau\right)=\dot{\phi}_{i}\left(\tau\right),
\end{equation}
noting that $\phi\left(\tau\right)$ is also periodic:
\begin{equation}
\phi_{i}\left(\beta\right)=\phi_{i}\left(0\right).
\end{equation}
The partition function in Eq. (\ref{eq:partf}) is now split into
three terms:
\begin{align}
\mathcal{Z}= & \int\left[\mathcal{D}\bar{a}\mathcal{D}a\right]e^{-\mathcal{S}_{1}\left[\bar{a},a\right]}\int\frac{dV^{S}}{2\pi}e^{-\mathcal{S}_{2}\left[n,V^{S}\right]}\times\nonumber \\
 & \int\mathcal{D}\phi\,e^{-\mathcal{S}_{3}\left[n,\dot{\phi}\right]},
\end{align}
where $\mathcal{S}_{1}$ remains unchanged as in Eq.(\ref{eq:s1})
and
\begin{align}
\mathcal{S}_{2}= & \beta\sum_{i}\Biggl[\frac{1}{2U}\left(V_{i}^{S}\right)^{2}+\nonumber \\
 & +\int_{0}^{\beta}d\tau\,\left(-\frac{\tilde{\mu}^{2}}{2U}-\frac{\tilde{\mu}}{iU}V_{i}^{S}-\frac{iV_{i}^{S}}{\beta}n_{i}\left(\tau\right)\right)\Biggr],\\
\mathcal{S}_{3}= & \sum_{i}\int_{0}^{\beta}d\tau\,\Biggl[\frac{1}{2U}\left(\dot{\phi}_{i}\left(\tau\right)\right)^{2}+\nonumber \\
 & -\frac{\tilde{\mu}}{iU}\dot{\phi}_{i}\left(\tau\right)-i\dot{\phi}_{i}\left(\tau\right)n_{i}\left(\tau\right)\Biggr].
\end{align}

The next step is a local gauge transformation:
\begin{align}
a_{i}\left(\tau\right) & =e^{i\phi_{i}\left(\tau\right)}b_{i}\left(\tau\right),\\
\bar{a}_{i}\left(\tau\right) & =e^{-i\phi_{i}\left(\tau\right)}\bar{b}_{i}\left(\tau\right),
\end{align}
which must also be applied to the chemical potential, as defined in
Eq. (\ref{eq:chempot}). This transformation, combined with the parametrization
$b_{i}\left(\tau\right)=b_{0}+b_{i}^{'}\left(\tau\right)$ we carry
out later on, reduces $\mathcal{S}_{2}$ entirely to a constant, so
it can be ignored in the path integral formulation. This leaves us
with
\begin{equation}
\mathcal{Z}=\int\left[\mathcal{D}\bar{b}\mathcal{D}b\right]\int\mathcal{D}\phi\,e^{-\mathcal{S}_{1}\left[\bar{b},b\right]}e^{-\mathcal{S}_{3}\left[n,\dot{\phi}\right]},\label{eq:partf3}
\end{equation}
the effective action terms now being
\begin{align}
\mathcal{S}_{1} & =\int_{0}^{\beta}d\tau\,\sum_{\left\langle i,j\right\rangle }\left[\bar{b}_{i}\left(\tau\right)g_{ij}^{1}b_{j}\left(\tau\right)+g_{ij}^{2}\left(\bar{b}_{i}\left(\tau\right)b_{j}\left(\tau\right)\right)^{2}\right],\label{eq:s1-1}\\
\mathcal{S}_{3} & =\sum_{i}\int_{0}^{\beta}d\tau\,\left[\frac{1}{2U}\left(\dot{\phi}_{i}\left(\tau\right)\right)^{2}-\frac{\tilde{\mu}}{iU}\dot{\phi}_{i}\left(\tau\right)\right],
\end{align}
where
\begin{align}
g_{ij}^{1} & =\delta_{ij}\frac{\partial}{\partial\tau}-Je^{-i\phi_{ij}\left(\tau\right)}-\frac{4\beta\bar{\mu}}{U}Te^{-i\phi_{ij}\left(\tau\right)},\\
g_{ij}^{2} & =-\frac{8\beta}{U}T^{2}e^{-i2\phi_{ij}\left(\tau\right)}.\label{eq:g2}
\end{align}

Here we have denoted $\phi_{ij}\left(\tau\right)=\phi_{i}\left(\tau\right)-\phi_{j}\left(\tau\right)$.
The similarity to an extended Quantum Phase Model (QPM) Hamiltonian
can already be seen at this point in the presence of both $e^{-i\phi_{ij}\left(\tau\right)}$-
and $e^{-i2\phi_{ij}\left(\tau\right)}$- dependent terms, which correspond
to cosine and double cosine parts of the action. The cosine expression
can be found in the QPM and describes the superfluid phase. The double
cosine term, then, must correspond to condensation of bosonic pairs.
Therefore we can clearly see from Eq. (\ref{eq:g2}) the impact of
the additional term Eq. (\ref{eq:density}) on the original bosonic
system. Due to $J$ having been defined as $J=t-2T$, our cosine term
contains two parts dependent on $T$:
\begin{equation}
+2Te^{-i\phi_{ij}\left(\tau\right)}-\frac{4\beta\bar{\mu}}{U}Te^{-i\phi_{ij}\left(\tau\right)}.
\end{equation}
The first part reduces the bosonic condensation with an amplitude
$2T$, irrespectively of the temperature and densities. The second
term competes with the first, strengthening the superfluid phase in
regions of higher densities and low temperatures. This can come as
a surprise in comparison with the effective model some naively assume,
which consists of two independent parts:
\begin{equation}
\mathcal{H}=J_{1}\sum_{\left\langle i,j\right\rangle }\cos\left(\phi_{ij}\right)+J_{2}\sum_{\left\langle i,j\right\rangle }\cos\left(2\phi_{ij}\right).
\end{equation}
To maintain physical clearness and integrity, the coefficients in
this model cannot be assumed and must be rigorously derived. As it
turns out, $J_{1}$ and $J_{2}$ are not constant and might also be
temperature dependent, as we show later. 

Furthermore we notice also the pair condensation term, which can lead
to pair condensation. Its dependence is proportional to $\thicksim T^{2}$,
rather than a linear dependence, as that of $\thicksim t$ in the
cosine term. 

To sum this part up, we emphasize that apart from pair condensation,
we distinguish two contrasting effects on the superfluid phase that
stem from the density induced term. In the whole range of temperatures
the DIT tends to have a dissipative influence on the original bosonic
system, but the situation can be different for higher densities and
low temperatures, where it works in favor of the superfluid phase.
Up to now, all calculations have been exact and the phenomena we analyze
stem from the density induced term. The assumptions made in order
to obtain the phase diagram we discuss in the next paragraph.

\subsubsection{Matrix form of effective action}

Before we go further with calculations, we must concentrate on the
regions we are interested in and physical phenomena we would like
to describe. We do not focus on the lob-like phase diagram, which
has already been established in the mean field approximation and which
would have to be calculated in a different way. Instead, we would
like to explore the specific heat (CH) of the system and ask the question
whether a second $\lambda$-like peak appears therein that would provide
clear proof of a second phase transition: in our case, the condensation
of bosonic pairs. Because the CH measures energy fluctuations, it
provides useful information about the system we analyze. From now
on, we make the necessary assumptions and explain what information
might be lost due to those assumptions.

The next step in order to achieve a phase-only model is getting rid
of $b_{i}$ by carrying out the following integral:
\begin{equation}
\int\left[\mathcal{D}\bar{b}_{i}\mathcal{D}b_{i}\right]\,e^{-\mathcal{S}_{1}\left[\bar{b},b\right]}.\label{eq:rid}
\end{equation}
For this to be possible, $\mathcal{S}_{1}$ in Eq. (\ref{eq:s1-1})
must be quadratic in bosonic field variables. The quadruple term is
split using a Wick average:
\begin{align}
\sum_{\left\langle i,j\right\rangle }b_{i}^{\dagger}b_{i}^{\dagger}b_{j}b_{j}\simeq & \sum_{\left\langle i,j\right\rangle }\Biggl[\left\langle b_{i}b_{i}\right\rangle b_{i}^{\dagger}b_{i}^{\dagger}+\left\langle b_{i}^{\dagger}b_{i}^{\dagger}\right\rangle b_{j}b_{j}+\nonumber \\
 & +\left(4\left\langle b_{i}^{\dagger}b_{j}\right\rangle +\delta_{ij}\right)b_{i}^{\dagger}b_{j}\Biggr],
\end{align}
which in our case gives
\begin{align}
\mathcal{S}_{1}= & \int_{0}^{\beta}d\tau\,\Biggl[\sum_{\left\langle i,j\right\rangle }\bar{b}_{i}\left(\tau\right)g_{ij}^{1}b_{j}\left(\tau\right)+\nonumber \\
 & +\sum_{\left\langle i,j\right\rangle }g_{ij}^{2}\left(\left\langle b_{j}b_{j}\right\rangle \bar{b}_{i}\bar{b}_{i}+\left\langle \bar{b}_{i}\bar{b}_{i}\right\rangle b_{j}b_{j}\right)\Biggr],
\end{align}
where 
\begin{align}
g_{ij}^{1}= & \,\delta_{ij}\frac{\partial}{\partial\tau}-Je^{-i\phi_{ij}\left(\tau\right)}-\frac{4\bar{\mu}}{U}Te^{-i\phi_{ij}\left(\tau\right)}+\nonumber \\
 & -\frac{8}{U}T^{2}e^{-i2\phi_{ij}\left(\tau\right)}\cdot\left(4\left\langle \bar{b}_{i}b_{j}\right\rangle +\delta_{ij}\right),
\end{align}
and $g_{ij}^{2}$ remains unchanged as in Eq. (\ref{eq:g2}). This
part is rather formal; nonlocal interactions are excluded in the process.
We rewrite $\mathcal{S}_{1}$ in matrix form, expanding the usual
one- or two-dimensional description of the Bose Hubbard model by introducing
a four-dimensional Nambu-like space:
\begin{align}
\mathcal{S}_{1} & =\boldsymbol{\bar{B}}\Gamma\boldsymbol{B},
\end{align}
where the vectors consist of bosonic fields,
\begin{equation}
\boldsymbol{\bar{B}}=\left(\begin{array}{cccc}
\bar{b}_{i} & b_{j} & \bar{b}_{j} & b_{j}\end{array}\right),
\end{equation}
\begin{equation}
\boldsymbol{B}=\left(\begin{array}{c}
b_{i}\\
\bar{b}_{i}\\
b_{j}\\
\bar{b}_{j}
\end{array}\right),
\end{equation}
and the matrix itself takes the form
\begin{equation}
\Gamma=\left(\begin{array}{cccc}
0 & \frac{1}{2}\delta_{ij}\Delta_{i} & \frac{1}{2}S_{ij} & 0\\
\frac{1}{2}\delta_{ij}\bar{\Delta}_{i} & 0 & 0 & 0\\
0 & 0 & 0 & \frac{1}{2}\delta_{ij}\Delta_{i}\\
0 & \frac{1}{2}S_{ij} & \frac{1}{2}\delta_{ij}\bar{\Delta}_{i} & 0
\end{array}\right),
\end{equation}
with
\begin{align}
S_{ij}= & \delta_{ij}\frac{\partial}{\partial\tau}-Je^{-i\phi_{ij}\left(\tau\right)}-\frac{4\bar{\mu}}{U}Te^{-i\phi_{ij}\left(\tau\right)}+\nonumber \\
 & -\frac{8}{U}T^{2}e^{-i2\phi_{ij}\left(\tau\right)}\cdot\left(4\left\langle \bar{b}_{i}b_{j}\right\rangle +\delta_{ij}\right),\\
\Delta_{i}= & \frac{8}{U}T^{2}e^{-i2\phi_{ij}\left(\tau\right)}\left\langle b_{i}b_{i}\right\rangle ,\\
\bar{\Delta}_{i}= & \frac{8}{U}T^{2}e^{-i2\phi_{ij}\left(\tau\right)}\left\langle \bar{b}_{i}\bar{b}_{i}\right\rangle .
\end{align}
After analytically diagonalizing $\Gamma$, the non-phase field dependent
part of the partition function, Eq. (\ref{eq:rid}), is now a Gaussian
integral,
\begin{equation}
\int\left[\mathcal{D}\bar{b}_{i}'\mathcal{D}b_{i}'\mathcal{D}\bar{b}_{j}'\mathcal{D}b_{j}'\right]\,e^{-\int_{0}^{\beta}d\tau\,\bar{B}\Gamma'B}=\det\Gamma'=e^{\textrm{Tr}\ln\Gamma'^{-1}},
\end{equation}
where $\Gamma'$ is the diagonalised matrix,
\begin{equation}
\Gamma'=\left(\begin{array}{cccc}
-\lambda_{1} & 0 & 0 & 0\\
0 & \lambda_{1} & 0 & 0\\
0 & 0 & -\lambda_{2} & 0\\
0 & 0 & 0 & \lambda_{2}
\end{array}\right),
\end{equation}
with eigenvalues
\begin{align}
\lambda_{1} & =\frac{1}{2}\sqrt{\bar{\Delta}_{i}\Delta_{i}-S_{ij}\sqrt{\bar{\Delta}_{i}\Delta_{i}}},\\
\lambda_{2} & =\frac{1}{2}\sqrt{\bar{\Delta}_{i}\Delta_{i}+S_{ij}\sqrt{\bar{\Delta}_{i}\Delta_{i}}}.
\end{align}
The entire partition function from Eq. (\ref{eq:partf3}) can be written
in the form
\begin{equation}
\mathcal{Z}=\int\mathcal{D}\phi\,e^{-\sum_{i}\int_{0}^{\beta}d\tau\,\left[\frac{1}{2U}\left(\dot{\phi}_{i}\left(\tau\right)\right)^{2}-\frac{\tilde{\mu}}{iU}\dot{\phi}_{i}\left(\tau\right)\right]}\cdot e^{\textrm{Tr}\ln\Gamma'}.
\end{equation}
We approximate, as usual, the trace of $\Gamma'$, to have quadratic
terms in the action only
\begin{align}
\mathrm{Tr}\ln\Gamma' & \approx\ln\left(\bar{\Delta}_{i}\Delta_{i}-S_{ij}^{2}\right)\approx\nonumber \\
 & \approx\ln\left(G_{0}^{-1}\right)^{2}+G_{0}^{2}\left[\bar{\Delta}_{i}\Delta_{i}-\left(S_{ij}'\right)^{2}\right]+2S_{ij}'G_{0},
\end{align}
where now
\begin{align}
S'_{ij}= & -Je^{-i\phi_{ij}\left(\tau\right)}+\frac{4\bar{\mu}}{U}Te^{-i\phi_{ij}\left(\tau\right)}+\nonumber \\
 & +\frac{8}{U}T^{2}e^{-i2\phi_{ij}\left(\tau\right)}\cdot\left(4\left\langle \bar{b}_{i}b_{j}\right\rangle +\delta_{ij}\right).
\end{align}
We parametrize the boson fields, $b_{i}\left(\tau\right)=b_{0}+b_{i}^{'}\left(\tau\right)$,
assuming any fluctuations are contained in the phase and fixing the
amplitude at a constant value. This approach can be very successful
when the dynamics of a system depend both on the amplitude and phase.
The coherence of the latter provides the phase transition between
ordered (superfluid) and disordered (normal insulator) phase. Thus,
$G_{0}=b_{0}^{2}$ can be calculated by minimizing the Hamiltonian,
$\partial H\left(b_{0}\right)/\partial b_{0}=0$, giving 
\begin{equation}
b_{0}^{2}=\frac{z\left(t-4T\right)+\left(\frac{U}{2}+\mu\right)}{U-8zT},
\end{equation}
which finally brings us to the final form of 
\begin{align}
\textrm{Tr}\ln\Gamma'= & \int_{0}^{\beta}d\tau\,\sum_{\left\langle i,j\right\rangle }\ensuremath{\left\{ b_{0}^{4}\left[\bar{\Delta}_{i}\Delta_{i}-\left(S_{ij}'\right)^{2}\right]+2S_{ij}'b_{0}^{2}\right\} }.
\end{align}
All that is left in this step is calculating $\left\langle \bar{b}_{i}b_{j}\right\rangle $
and the anomalous averages $\left\langle \bar{b}_{i}\bar{b}_{i}\right\rangle $
and $\left\langle b_{i}b_{i}\right\rangle $. The anomalous averages
can be rewritten as
\begin{equation}
\left\langle b_{i}b_{i}\right\rangle =b_{0}^{2}\left\langle e^{i2\phi_{i}}\right\rangle =b_{0}^{2}\Psi_{2\phi},
\end{equation}
\begin{equation}
\left\langle \bar{b}_{i}\bar{b}_{i}\right\rangle =b_{0}^{2}\left\langle e^{-i2\phi_{i}}\right\rangle =b_{0}^{2}\Psi_{2\phi},
\end{equation}
where $\Psi_{2\phi}=\left\langle e^{i2\phi}\right\rangle =\left\langle e^{-i2\phi_{i}}\right\rangle $
is the pair condensation order parameter, in which we neglect the
chirality of the phase. The average $\left\langle \bar{b}_{i}b_{j}\right\rangle $
is equal to
\begin{align}
\left\langle \bar{b}_{i}b_{j}\right\rangle  & =b_{0}^{2}\left\langle e^{-i\left[\phi_{i}\left(\tau\right)-\phi_{j}\left(\tau'\right)\right]}\right\rangle =b_{0}^{2}\,G_{ij}\left(\tau,\tau'\right),
\end{align}
where
\begin{align}
G_{ij}\left(\tau,\tau'\right)= & \delta_{ij}e^{\frac{U}{2}\left|\tau-\tau'\right|}\cdot\gamma_{ij}\left(\tau,\tau'\right)
\end{align}
is the Green's function {[}\citealp{refqr}{]}, with
\begin{align}
 & \gamma_{ij}\left(\tau,\tau'\right)=\nonumber \\
 & \frac{\sum_{n_{i}}\exp\left[-\frac{U\beta}{2}\left(n_{i}+\frac{\bar{\mu}}{U}\right)^{2}\right]\exp\left[-U\left(n_{i}+\frac{\bar{\mu}}{U}\right)\left(\tau-\tau'\right)\right]}{\sum_{n_{i}}\exp\left[-\frac{U\beta}{2}\left(n_{i}+\frac{\bar{\mu}}{U}\right)^{2}\right]}\\
 & \approx\frac{\coth\left[\frac{\beta}{2}\left(\frac{U}{2}-\bar{\mu}\right)\right]+\coth\left[\frac{\beta}{2}\left(\frac{U}{2}+\bar{\mu}\right)\right]}{2}.
\end{align}
After these operations, the final form of the partition function,
barring constant terms and (as a second-order approximation) quadrupolar
phase exponent terms, is
\begin{align}
\mathcal{Z}= & \int\mathcal{D}\phi\,\Biggl[e^{-\sum_{i}\int_{0}^{\beta}d\tau\,\left[\frac{1}{2U}\left(\dot{\phi}_{i}\left(\tau\right)\right)^{2}-\frac{\tilde{\mu}}{iU}\dot{\phi}_{i}\left(\tau\right)\right]}\nonumber \\
 & \times e^{\sum_{\left\langle i,j\right\rangle }\int_{0}^{\beta}d\tau\,\left(\varepsilon_{1}e^{-i2\phi_{ij}\left(\tau\right)}+\varepsilon_{1}e^{-i\phi_{ij}\left(\tau\right)}\right)}\Biggr],\label{eq:partf2tolast}
\end{align}
where
\begin{align}
\varepsilon_{1}= & \left[\frac{z\left(t-4T\right)+\bar{\mu}}{U-8zT}\right]^{2}\left[\frac{64\bar{\mu}}{U^{2}}T^{3}-\frac{16}{U}\left(t-2T\right)T^{2}\right]\\
 & \times\left\{ 2\left[\coth\frac{\beta\mu}{2}+\coth\frac{\beta\left(\mu+U\right)}{2}\right]+1\right\} +\\
 & +\frac{z\left(t-4T\right)+\bar{\mu}}{U-8zT}\left[\frac{8\bar{\mu}}{U}T-2\left(t-2T\right)\right],\\
\varepsilon_{2}= & \left[\frac{z\left(t-4T\right)+\bar{\mu}}{U-8zT}\right]^{2}\\
 & \times\left[\left(t-2T\right)^{2}+\left(\frac{4\bar{\mu}}{U}T\right)^{2}-2\left(t-2T\right)\frac{8\bar{\mu}}{U}T\right].
\end{align}
We see clearly now that the already mentioned naive past assumptions
about constant values of the amplitudes in this phase model have no
justification in reality. The coefficients $\varepsilon_{1}$ and
$\varepsilon_{2}$ have complex structures, even though we dropped
the lattice dependence, leaving in only the coordination number $z$.
We also note that $\varepsilon_{2}$, which comes from the DIT term,
is temperature dependent.

\subsection{Transformation to $S=1$ pseudospin}

Assuming the on-site two-particle interaction is strong, which is
a reasonable condition for this model, we can ignore the complex term
in Eq. (\ref{eq:partf2tolast}), getting
\begin{align}
\mathcal{Z}= & \int\mathcal{D}\phi\,\Biggl\{ e^{-\int_{0}^{\beta}d\tau\,\sum_{i}\frac{1}{2U}\left(\dot{\phi}_{i}\left(\tau\right)\right)^{2}}\times\nonumber \\
 & e^{-\int_{0}^{\beta}d\tau\,\left[-\sum_{\left\langle i,j\right\rangle }\left(\varepsilon_{2}e^{-i2\phi_{ij}\left(\tau\right)}+\varepsilon_{1}e^{-i\phi_{ij}\left(\tau\right)}\right)\right]}\Biggr\}.
\end{align}
This simplification excludes the accurate description of the properties
of the system with chemical potential variation. The partition function
corresponds to the following phase hamiltonian: 
\begin{align}
\hat{H}= & -4U\sum_{i}\left(\frac{1}{i}\frac{\partial}{\partial\hat{\phi}_{i}}\right)^{2}-\sum_{\left\langle i,j\right\rangle }\varepsilon_{1}\cos\left(\hat{\phi}_{i}-\hat{\phi}_{j}\right)+\nonumber \\
 & -\sum_{\left\langle i,j\right\rangle }\varepsilon_{2}\cos\left[2\left(\hat{\phi}_{i}-\hat{\phi}_{j}\right)\right].
\end{align}
The two interaction terms give rise to two different ordered phases,
represented by two order parameters:
\begin{align}
\Psi_{\phi} & \equiv\left\langle e^{i\phi}\right\rangle ,\label{eq:op1-1}\\
\Psi_{2\phi} & \equiv\left\langle e^{i2\phi}\right\rangle .\label{eq:op2-1}
\end{align}
$\Psi_{\phi}$ is the superfluid order parameter, known from the pure
BH model; $\Psi_{2\phi}$ corresponds to the phenomenon of bosonic
pair tunneling.

\subsubsection{Pure Bose-Hubbard mapping}

The matrix elements of the phase operator in its own basis are 
\begin{equation}
\left\langle k\left|N\left(\phi\right)\right|m\right\rangle =\int_{0}^{2\pi}\frac{d\phi}{2\pi}\,e^{-ik\phi}\left(\frac{1}{i}\frac{\partial}{\partial\phi}\right)e^{im\phi}=m\delta_{k,m}.\label{eq:n-1}
\end{equation}
The other operators needed can be derived from Eq. (\ref{eq:n-1})
{[}\citealp{simanek}{]}, giving
\begin{align}
\left\langle k\left|\cos\phi\right|m\right\rangle  & =\int_{0}^{2\pi}\frac{d\phi}{2\pi}\,e^{-i\left(k-m\right)\phi}\cos\phi=\nonumber \\
 & =\frac{1}{2}\left(\delta_{k-m-1,0}+\delta_{k-m+1,0}\right),\\
\left\langle k\left|\sin\phi\right|m\right\rangle  & =\frac{i}{2}\left(\delta_{k-m-1,0}-\delta_{k-m+1,0}\right).
\end{align}
For spin $S=1$, $k,m$ are limited to the lowest-energy states: $-1,0,1$.
We have assumed that $U\rightarrow\infty$, which in particular means
that $k_{B}T/U$ is small, and
\begin{eqnarray}
N\left(\phi\right) & = & S_{z},\\
\cos\phi & = & \frac{1}{\sqrt{2}}S_{x},\\
\sin\phi & = & \frac{1}{\sqrt{2}}S_{y}.
\end{eqnarray}
 First, we only transform the first two terms of the Hamiltonian to:
\begin{align}
\mathcal{H} & =U\sum_{i}N^{2}-\sum_{\left\langle i,j\right\rangle }\varepsilon_{1}\cos\left(\phi_{i}-\phi_{j}\right)=\nonumber \\
 & =U\sum_{i}\left(S_{i}^{z}\right)^{2}-\frac{1}{2}\varepsilon_{1}\sum_{\left\langle i,j\right\rangle }\left(S_{i}^{x}S_{j}^{x}+S_{i}^{y}S_{j}^{y}\right).
\end{align}
 If $T=0$, we have at this point a model analogous to the pure BH
model, as well as to the QPM Hamiltonian. Applying a mean field approximation:
$\left\langle S_{i}^{y}\right\rangle =0$;
\begin{equation}
S_{i}^{x}S_{j}^{x}\approx\left\langle S_{i}^{x}\right\rangle S_{j}^{x}+S_{i}^{x}\left\langle S_{j}^{x}\right\rangle -\left\langle S_{i}^{x}\right\rangle \left\langle S_{j}^{x}\right\rangle ,
\end{equation}
 we arrive at the following Hamiltonian:

\begin{align}
\mathcal{H} & =U\left(S_{i}^{z}\right)^{2}-\frac{1}{2}\varepsilon_{1}S_{i}^{x}\left\langle S_{i}^{x}\right\rangle =\nonumber \\
 & =J\left[\frac{U}{J}\left(S_{i}^{z}\right)^{2}-S_{i}^{x}\Psi_{e}\right],
\end{align}
where $J=\frac{1}{2}z\varepsilon_{1}$ and $\Psi_{\phi}$ is the superfluid
order parameter.

\subsubsection{Adding the double interaction}

We define the bilinear superexchange terms:
\begin{eqnarray}
Q_{i} & = & \left(S_{i}^{x}\right)^{2}-\left(S_{i}^{y}\right)^{2},\\
Q_{i}^{xy} & = & 2S_{i}^{x}S_{i}^{y},
\end{eqnarray}
and perform a mean field approximation
\begin{equation}
Q_{i}Q_{j}\approx\left\langle Q_{i}\right\rangle Q_{j}+Q_{i}\left\langle Q_{j}\right\rangle -\left\langle Q_{i}\right\rangle \left\langle Q_{j}\right\rangle .
\end{equation}
Assuming $\left\langle Q_{i}^{xy}\right\rangle =0$, the full mean
field pseudospin Hamiltonian is 
\begin{align}
\mathcal{H} & =U\left(S_{i}^{z}\right)^{2}-\frac{1}{2}z\varepsilon_{1}S_{i}^{x}\left\langle S_{i}^{x}\right\rangle -\frac{1}{4}z\varepsilon_{2}\left(Q_{i}\left\langle Q_{i}\right\rangle \right)=\nonumber \\
 & =J\left[\frac{U}{J}\left(S_{i}^{z}\right)^{2}-S_{i}^{x}\Psi_{\phi}-\frac{J_{2}}{J}Q_{i}\Psi_{2\phi}\right],
\end{align}
where
\begin{align}
J & =\frac{1}{2}z\varepsilon_{1},\\
J_{2} & =\frac{1}{4}z\varepsilon_{2},
\end{align}
We define the system's free energy per site as {[}\citealp{refkopec}{]}
\begin{equation}
f=\frac{1}{2}\left(J\Psi_{\phi}^{2}+J_{\perp}\Psi_{2\phi}^{2}\right)-\frac{1}{\beta}\ln Z.
\end{equation}
The two order parameters, Eq. (\ref{eq:op1-1}) and Eq. (\ref{eq:op2-1})
then minimize the free energy, and their values can be calculated
from the following self-consistent equations:

\begin{eqnarray}
\frac{\partial f}{\partial\Psi_{\phi}}=0, & \quad & \frac{\partial f}{\partial\Psi_{2\phi}}=0,
\end{eqnarray}
which in this case are

\begin{equation}
1=\frac{4J\tanh\left[\beta/2\sqrt{\left(U-J_{2}\Psi_{2\phi}\right)^{2}+4J^{2}\Psi_{\phi}^{2}}\right]}{\sqrt{\left(U-J_{2}\Psi_{2\phi}\right)^{2}+4J^{2}\Psi_{\phi}^{2}}\left[X+2\right]},\label{eq:sc1-1}
\end{equation}
\begin{equation}
\Psi_{2\phi}=\frac{U}{J_{2}-4J}+\frac{4J}{4J-J_{2}}\cdot\frac{1-X}{2+X},\label{eq:sc2-1}
\end{equation}
where
\begin{equation}
X=\frac{e^{-\frac{\beta}{2}\left(U+3J_{2}\Psi_{2\phi}\right)}}{\cosh\left[\beta/2\sqrt{\left(U-J_{2}\Psi_{2\phi}\right)^{2}+4J^{2}\Psi_{\phi}^{2}}\right]}.
\end{equation}
The critical line equations in Eqs. (\ref{eq:sc1-1}) and (\ref{eq:sc2-1})
allow us to obtain phase diagrams for any chosen parameters of the
on-site interaction $U$, the chemical potential $\mu$, the temperature
$T_{C}$ (so labelled to avoid confusion with the density-induced
tunneling parameter), the pure BH hopping $t$ and the density-induced
tunneling amplitude $T$.

\section{Results}

\begin{widetext}
\begin{figure}
\includegraphics[scale=0.8]{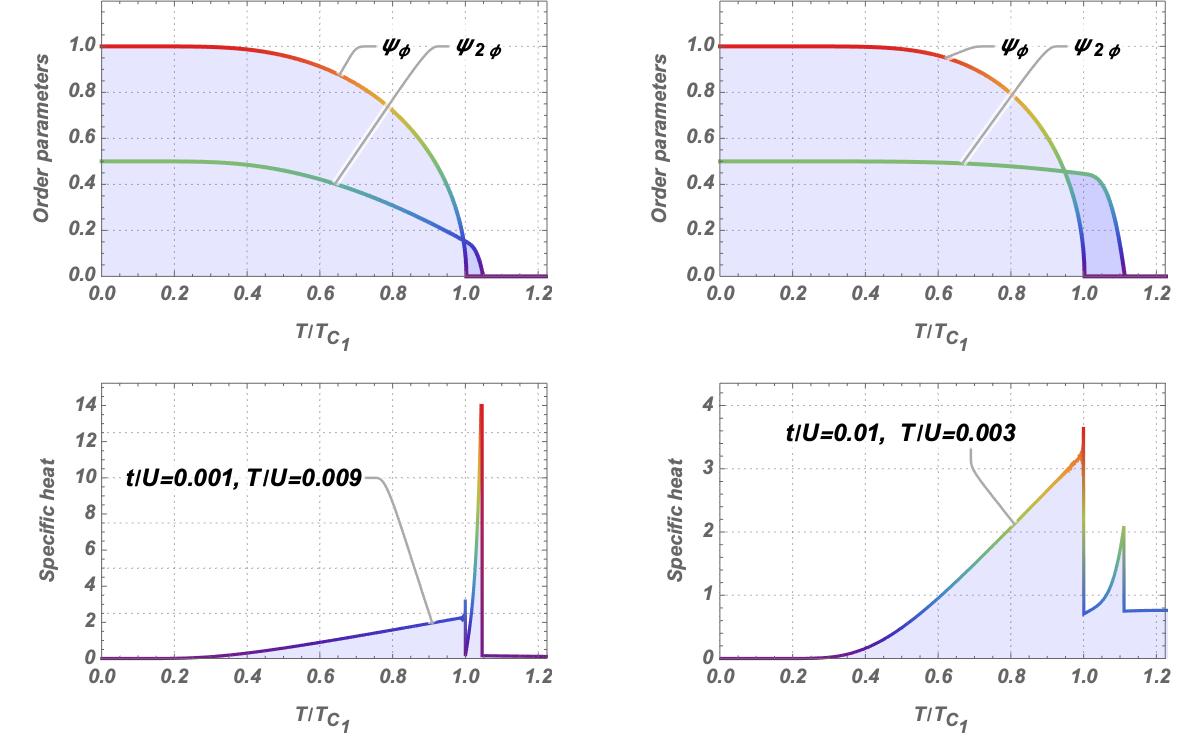}\caption{Upper row: temperature dependence of single $\Psi_{\phi}$ and pair
$\Psi_{2\phi}$ order parameters. Bottom row the specific heat of
the system for chosen parameters ($\mu/U=1.42$).}
\label{fig1}
\end{figure}

\end{widetext}

Below are some exemplary diagrams obtained with use of Eqs. (\ref{eq:sc1-1})
and (\ref{eq:sc2-1}). First of all, Fig. (\ref{fig1}) shows the
dependence of the single and pair order parameters on the normalized
temperature $T/T_{C_{1}}$. The normalization is taken as $T_{C_{1}}$,
which is the critical temperature connected to the single bosonic
condensation phase transition, which separates the single $\Psi_{\phi}$
and pair $\Psi_{2\phi}$ superfluid phase. We have chosen parameter
values for which phase separation can be clearly seen. This is the
most interesting observation we have made so far: not only are there
two separate, coexisting superfluid phases in this model; pair condensation
also occurs independently of single-particle condensation. We can
also see that even though a mean field approximation was used in the
later stages of pseudospin mapping, the system retained enough information
that we were able to expose phenomena that eluded mean-field-only-based
approaches. Interestingly, the pair condensation survives at higher
temperatures than single bosonic condensation, even as we change the
density of the particles and the energy scales. In the range of parameters
where $t/T<1$ (pair energy scales are higher) we see that the single
particle condensation is almost suppressed and energy fluctuations
are enormous, but pretty narrow in the temperature range. This is
contrary to the opposite case, when $t/T>1$, where one can see a
strong single superfluid phase and a well established and separated
pair condensed fraction. We note that there is no region with only
$\Psi_{\phi}\neq0$ and the phase transitions are lambda-like, already
observed experimentally. 

\begin{figure}
\includegraphics[scale=0.7]{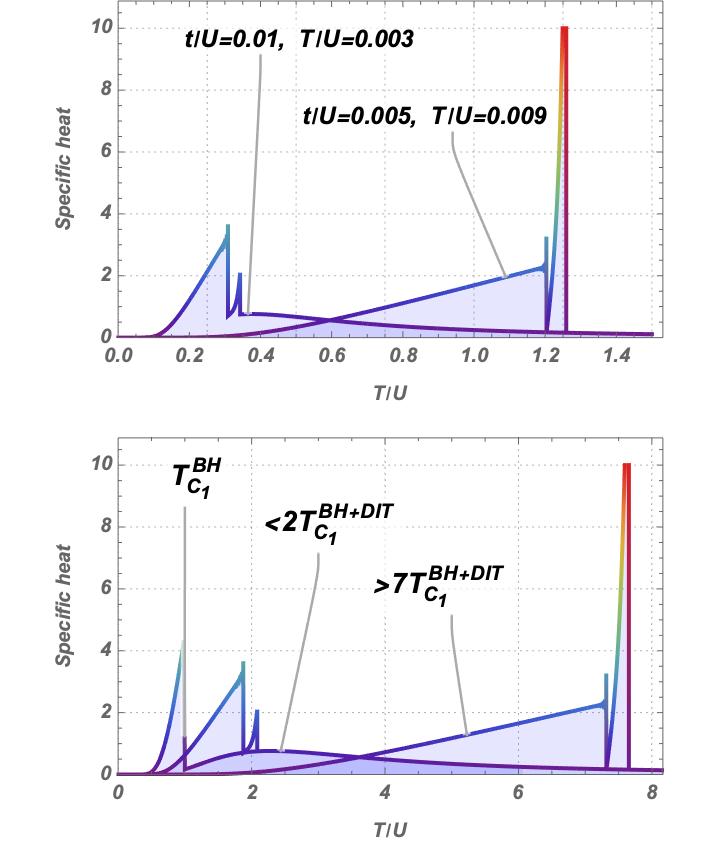}\caption{Upper diagram: comparison of the specific heat versus temperature
dependence for the opposite choice of the single and pair energy scales
$t/T>1$ left peaks and $T/t<1$ - right respectively. Bottom diagram:
impact of the density induced therm on single boson condensation $T_{C_{1}}$
for the choice parameter from left plot ($\mu/U=1.42$).}
\label{fig2}
\end{figure}

Although normalization was taken to clarify the amplitude of the energy
calculations, we now move forward without it to observe the actual
temperature dependence of the thermodynamic function Fig. (\ref{fig2}).
What occurs is an interesting phenomenon. Although higher values of
DIT energy give rise to higher critical temperatures of the single
condensation $T_{C_{1}}$, it simultaneously suppresses the superfluid
phase, providing a strong response in the pair sector. On the other
hand, in the opposite regime, the pair superfluid phase ceases to
exist, providing support for the pure BH model superfluidity with
an increase in value of the critical temperature $T_{C_{1}}/U$. If
we take the value of the DIT equal $T/U=0.009$, the critical temperature
$T_{C_{1}}$ of the single particle condensation becomes approximately
seven times larger; for $T/U=0.003$, it is almost twice as large. 

\section{Summary}

In this work we have presented an analytical study of the density-induced
tunneling Bose-Hubbard model. We utilized methods known for their
high accuracy in order to receive a fuller picture than mean field
theory could provide, considering the model within a path integral
formulation of quantum mechanics and applying the U(1) quantum rotor
method. Those methods allowed us to rewrite the effective action,
and, by extension, the Hamiltonian, as phase-only, to map it onto
a $S=1$ pseudospin model and from that obtain critical line equations.

Thanks to the quantum rotor method, which has proved its accuracy
in other systems, and especially its preservation of multi-particle
correlations, we have managed to shed light on the existence of a
previously unconfirmed pair superfluid phase at finite temperatures
in the density-induced tunneling BH model. What's more, we have shown
that, for certain parameter values, this phase occurs exclusively
where single-particle condensation does not. Despite the complications
caused by DIT, we managed to obtain the specific heat and observe
regions where energy fluctuations are highest and (in accordance with
order parameters) accurately point out the phase transitions. These
phases we recognized as the usual Bose condensation and an additional,
previously unaccounted for, bosonic pair condensation at finite temperatures.
We conclude from our analysis that there are different ways in which
DIT impacts the pure BH system. For large values of the density-induced
amplitude, the critical temperature of single particle condensation
is higher and the specific heat has a sharp peak (well known lambda
behavior). For lower values of the tunneling amplitude (ie. less than
the tunneling amplitude for pure BH), the peak in the thermodynamic
function is broader. Of course, the results shown in this work call
for experimental confirmation, but, once confirmed, could potentially
introduce a new branch of thought in optical lattice-related research.

The analytical framework established in this paper can serve as a
foundation for the analysis of any number of properties of the density-induced
tunneling BH model, as well as its modifications, such as external
magnetic fields, particle mixtures or various lattice geometries beyond
the simple cubic lattice here considered. We plan to make use of this
framework in future research.
\begin{acknowledgments}
One of us (T.P.P.) would like to acknowledge that this work has been
done under the Maestro Grant No. DEC-2019/34/A/ST2/00081 of the Polish
National Science Centre (NCN).
\end{acknowledgments}

\bibliographystyle{apsrev4-1}
\nocite{*}
\bibliography{biblio}

\end{document}